# COMPARATIVE STUDY OF PERFORMANCE OF WIND WAVE MODEL: WAVEWATCH—MODIFIED BY NEW SOURCE FUNCTION

V. G. Polnikov* and V. Innocentini**

*Obukhov Institute for Physics of Atmosphere of the Russian Academy of Sciences,
Pyzhevkii Lane 3, 119017 Moscow, Russia
E-Mail: polnikov@mail.ru (Corresponding Author)
**Center of Weather Forecast and Climate Studies,
Av. Astronautos 1758, Sao Jose dos Campos, CEP 12227-010 SP, Brazil

**ABSTRACT:** With the aim of assessing the merits of the new source function proposed earlier in Polnikov (2005), it was tested and validated by means of the modification of the well known model WAVEWATCH-III. Assessment was done on the basis of comparing the wave simulation results found from both models against the buoy data obtained in three oceanic regions: eastern and western parts of the North Atlantic, and the Barenz Sea. First of all, incorporation of the new source function into the numerical codes of WAVEWATCH was done, and this modified version of WAVEWATCH was fitted and tested by standard tests. On the basis of these results some physical conclusions were drawn. Then, sophisticated fitting of the modified model was carried out, using the observation data of 19 buoys obtained in two regions of the North Atlantic for a period of 30 days at 1-hour time intervals. After this, the standard validation of both models was continued on the basis of the 8-month historical data, corresponding to 12 buoys located in the western part of the North Atlantic. Finally, comparative validation of the models was done against the wave observations obtained at one buoy in the Barenz Sea for a period of 3 years at 6-hour time intervals. Estimations of simulation accuracy were carried out for the three parameters of wind waves: significant wave height, $H_s$, peak wave period, $T_p$, and mean wave period, $T_m$. Comparative analysis of these estimations was carried out for the original and modified model WAVEWATCH. The advantage of the modified model was revealed, with an increase of simulation accuracy for $H_s$ by 20 to 50% for more than 70% of the buoys considered. Additionally, it was found that the speed of calculation was increased by 15%.

## 1. INTRODUCTION

All modern numerical models for wind waves are based on the solution of the evolution equation for a two-dimensional wave energy spectrum, $S(\sigma,\theta,\mathbf{x},t)$ (or a wave action one, $N(...) = S(...)/\sigma$), given in the space of wave frequency, $\sigma$, and wave propagation angles, $\theta$, and spread through the geographic coordinates, $\mathbf{x}$, and time, $t$. In general, this equation has the form

$$\frac{dS(\sigma,\theta,\mathbf{x},t)}{dt} = F(S,\mathbf{W},\mathbf{U}) = In + Nl - Dis \qquad (1)$$

Here, the left hand side is the full derivative of the spectrum in time, and the right hand side is the source function, $F$, depending on both the wave spectrum, $S$, and the external wave-making factors: local wind, $\mathbf{W}(\mathbf{x}, t)$, and local current, $\mathbf{U}(\mathbf{x}, t)$. Here we shall restrict ourselves to the case of no currents because of the absence of information on surface currents in the areas under consideration. Herewith, we have taken into account the fact that, under this simplifying assumption, the relative error of calculations would be less than 5% due to the small value of typical currents in oceans (0.1–0.3 m/s) with respect to phase velocity of fully developed wind waves (5–15 m/s). More detailed consideration needs a separate study.

The source function describes certain physical mechanisms included in the model, which are responsible for the wave spectrum evolution (Efimov and Polnikov, 1991; Komen et al, 1994). It is common to distinguish three terms in function $F$: the atmosphere-wave energy exchange mechanism, $In$; the energy conservative mechanism of nonlinear wave-wave interactions, $Nl$; and the wave energy loss mechanism, $Dis$, related to the wave breaking and interaction of waves with the turbulence of the upper water layer and bottom. Differences in representation of the source function terms mentioned above determine general differences between wave models. In particular, the models are classified into categories of generations, by ranging the parameterization for *Nl*-term (The SWAMP group, 1985). In principle, this classification could be extended, taking into account all source function terms (for examples, see Polnikov, 2005; Polnikov and Tkalich, 2006).

Differences in representation of the left hand side of the evolution equation (1) and the realization of its numerical solution are related to the mathematics of a wave model, which determines the specificity of a model as well. This specificity is related to domains of applicability of the models (i.e., accounting for the sphericity of the Earth, the wave refraction due to bottom or current inhomogeneity and so on). Here we do not consider these effects; we focus on the physics of models only.

Three models of the third generation: WAM (The WAMDI group, 1988), WAVEWATCH (Tolman and Chalikov, 1996), and SWAN (Booij, Ris and Holthuijen, 1999) are the most widely used in the world at present. The first two are used for global wave forecast in deep water. The third one represents an elaboration of the first model for the case of finite depth water. It is mainly used for regional forecast.

The models mentioned are rather well fitted against observations and give satisfactory results of calculations. However, they have been constructed on physical grounds that are more than 10 years old. Therefore, despite of continuous updating, they are obsolete to some extent, both in the aspect of substantiation of the source function terms and in the aspect of technical realization of the mathematics of the models. All these circumstances restrict the potential applicability of the models. The regularly reported new theoretical findings and continuous extension of the domain of models application dictate the necessity to construct a new, more modern model. First of all, it is related to the modification of the source function, *F*. One such modification was proposed in the recent paper by Polnikov (2005), where it was called "the optimized source function" (for details, see the original paper). Our attempt to incorporate this new source function into the model WAM gave very encouraging results (Polnikov et al., 2008). We have found that the errors of numerical simulations were decreased by 1.5–2 times, whilst the speed of calculation was enhanced by 25%.

In the present paper, we pose the task of estimating the merits of the new source function by incorporating it into the mathematical codes of the model WAVEWATCH-III (version 2.22) (WW hereafter), which is the most advanced at present (Tolman et al., 2002). This estimation will be done on the basis of comparison of the numerical simulations against the buoy measurements of wind waves obtained in three parts of the world oceans: eastern and western parts of the North Atlantic, and the Barenz Sea.

The layout of the paper is as follows. Section 2 describes briefly the kind of evolution equation used in WW and the main features of the new source function. Section 3 is devoted to the methodology of our investigations, paying special attention to regulations of the processes of testing and comparative validation of the models. Simulation results of two testing cases and physical analysis of them are presented in Section 4, while the extensive presentation of the validation results is given in Section 5. Conclusions of the study and prospects of future work are stated in Section 6.

## 2. MODIFIED WAVEWATCH

### 2.1 Main evolution equation

In WW, the evolution equation is written in spherical coordinates, ($\phi, \lambda$), for the time-spatial distribution of the two-dimensional wave action spectrum, $N(k, \theta, \phi, \lambda, t)$, which is related to the energy spectrum, $S \equiv S(k, \theta, \phi, \lambda, t)$, by the linear ratio, $N(k, \theta, ...) = S(k, \theta, ...)/\sigma(k)$. This evolution equation has the typical form (Komen et al., 1994):

$$\frac{\partial N}{\partial t} + \frac{1}{\cos\phi}(\frac{\partial}{\partial \phi}\dot{\phi} N \cos\phi + \frac{\partial}{\partial \lambda}\dot{\lambda} N) + \frac{\partial}{\partial \theta}\dot{\theta}_g N$$
$$= F(N) = Nl + In - Dis \,, \qquad (1a)$$

where

$$\dot{\phi} = \frac{c_g \sin\theta + U_\phi}{R}, \qquad (1b)$$

$$\dot{\lambda} = \frac{c_g \cos\theta + U_\lambda}{R}, \qquad (1c)$$

$$\dot{\theta}_g = \dot{\theta} - \frac{c_g \tan\phi \cos\theta}{R}. \qquad (1d)$$

Here, $\phi, \lambda$ are the latitude and longitude coordinates, respectively; $\theta$ is the angle of wave component propagation with respect to latitude; $k$ and $\sigma(k)$ are the wave number and wave frequency of the component under consideration, related to each other by the dispersion relation $\sigma(k) = [gk\,\text{th}(kD)]^{1/2}$, where *D* is the local water depth; $c_g = \partial\sigma(k)/\partial k$ is the group velocity of the wave component; $(U_\phi, U_\lambda)$ are the components of mean ocean currents; $g$ is the gravitational acceleration constant; and $R$ is the Earth's radius.

The "heart" of the model is the source function, $F$. In WW, as usual, it contains three types of evolution terms describing the wind wave physics: nonlinearity-term, $Nl$; input-term, $In$; and dissipation-term, $Dis$, which includes the deep and shallow water summands. Here we do not give the cumbersome description of the terms used in the original WW. They are thoroughly described in Tolman (2002). But, the modified terms are briefly represented below following Polnikov (2005), to the extent that they are the objects of investigation. Note that they are given in the energy spectrum representation, $F(S)$, which is related to the wave action representation, $F(N)$, by the relation $F(N) = F(S)/\sigma(k)$, to avoid any misunderstanding.

In the present study, we consider the case of deep water and absence of the currents, $\mathbf{U}(\mathbf{x},t) = 0$. For this reason, the refraction processes and the bottom dissipation terms are not considered.

## 2.2 Nl-term

In the new source function, an optimized version of the Discrete Interaction Approximation (DIA), instead of the ordinary DIA used in WW, is used for the parameterization of $Nl$. The optimization includes the following items:

(a) Fast version of DIA (FDIA) (Polnikov and Farina, 2002)

(b) New, more effective configuration for the four interacting waves (Polnikov, 2003).

It is well known that DIA contains only one summand instead of an infinite number of summands under the four-wave kinetic integral describing the nonlinear evolution mechanism. To realize this approximation, one sets some initial components $(k,\theta)$ of the reference wave and calculate components of the other three waves by a certain scheme, using the resonant conditions of the interacting waves (see the original paper by Hasselmann et al. (1985) for details).

Item (a) above means a special reconstruction of the calculation procedure for the interacting components of the other three waves, making it possible to avoid the cumbersome procedure of spectrum interpolation used in the original DIA.

Item (b) means a certain choice of the four-waves configuration, leading to the result of DIA which is the closest to the result of the exact calculation of the kinetic integral. Finally, these two improvements lead to an increase of the speed of calculations and a better correspondence of approximate calculations of the $Nl$-term to the exact numerical values of the latter (for numerical details, see the original papers).

Formally, the fast DIA is governed by the following factors.

(1) The computational frequency-angular grid, $\{\sigma_i, \theta_j\}$, defined typically as:

$$\sigma(i) = \sigma_0 e^{i-1} \quad (1 \le i \le N), \text{ and}$$
$$\theta(j) = -\pi + (j-1)\cdot\Delta\theta \quad (1 \le j \le M), \quad (2)$$

where $\sigma_0$, $e$, and $\Delta\theta$ are the grid parameters specified below.

(2) The reference wave component, $(\sigma, \theta)$, located at a certain current node of grid (2).

(3) The other three waves which have components located at nodes of the same grid, defined by the relations

$$\sigma_1 = \sigma e^{m1}, \quad \sigma_2 = \sigma e^{m2}, \quad \sigma_3 = \sigma e^{m3}, \quad (3a)$$

$$\Delta\theta_1 = n1\Delta\theta, \quad \Delta\theta_2 = n2\Delta\theta, \quad \Delta\theta_3 = n3\Delta\theta,$$
(where $\Delta\theta_i \equiv |\theta_i - \theta|$). (3b)

Thus, the optimal configuration of the four interacting waves is given by the set of integer numbers: $m1, m2, m3; n1, n2, n3$, which, in turn, are to be specially calculated depending on the grid parameters, $e$ and $\Delta\theta$ (Polnikov and Farina, 2002).

For the frequency-angle grid with parameters $e = 1.1$ and $\Delta\theta = \pi/12$, which are typical for the model WW, the most effective configuration is given by the following set of parameters (Polnikov, 2003).

$$m1=3, m2=3, m3=5; n1=n2=2, n3=3 \quad (4)$$

Finally, the $Nl$-term is calculated by the standard formulas, making the loops for the reference components, $(\sigma,\theta)$, arranged through grid (2). For completeness, these formulas are as follows:

$$Nl(\sigma,\theta) = Nl(\sigma_3,\theta_3) = I(\sigma_1,\theta_1,\sigma_2,\theta_2,\sigma_3,\theta_3,\sigma,\theta), \quad (5a)$$

$$Nl(\sigma_1,\theta_1) = Nl(\sigma_2,\theta_2) = -I(\sigma_1,\theta_1,\sigma_2,\theta_2,\sigma_3,\theta_3,\sigma,\theta), \quad (5b)$$

where $\quad I(...) = C_{nl} g^{-4} \sigma^{11} \left[ S_1 S_2 (S_3 + (\sigma_3/\sigma)^4 S) - S_3 S((\sigma_2/\sigma)^4 S_1 + (\sigma_1/\sigma)^4 S_2) \right]$ (6)

Here, $C_{nl}$ is the only fitting non-dimensional coefficient, and the notation $S_i \equiv S(\sigma_i, \theta_i)$ is used.

## 2.3 In-term

General representation of the input term, typically used in all modern models, has the form

$$In = \beta(\sigma, \theta, \mathbf{W}) \sigma S(\sigma, \theta) \qquad (7)$$

In the new source function, we use the following form of increment $\beta$

$$\beta = C_{in} \max\left\{-b_L, \left[0.04\left(\frac{u_*\sigma}{g}\right)^2 + 0.00544\frac{u_*\sigma}{g} + 0.000055\right]\cos(\theta - \theta_w) - 0.00031\right\} \qquad (8)$$

where $u_*$ is the friction velocity, and $\theta_w$ is the local wind direction.

In contrast to the cumbersome theoretical form for $\beta(\sigma, \theta, \mathbf{W})$ used in WW, parameterization (8) is based on the empirical data. The part of the mathematical form of $\beta$, after the comma inside the curly brackets of (8), was derived by Yan (1987). The main advantage of this form is its validity in a wide range of normalized frequencies: $0.5 < W_{10}\sigma/g < 75$, where $W_{10}$ is the local wind velocity at the reference height, $z = 10$ m. In addition to this form, we have introduced a special feature of the increment, the existence of a negative value: $\beta = -b_L$. This value corresponds to wave components propagating with a velocity greater than the properly directed projection of the wind velocity. Such a feature of $\beta$ is physically important, as was proved in numerical studies by Chalikov (Tolman and Chalikov, 1996). In the model WW, the value of $b_L$ is calculated numerically by some formulas; but in our case, it is simply found by the fitting process.

In parameterization (8), the coefficient $C_{in}$ and parameter $b_L$ are the subjects of the model fitting. As it was found earlier (Polnikov, 2005), the default value for the latter is $b_L = 5 \cdot 10^{-6}$.

In this work, a transition $W_{10} \Leftrightarrow u_*$ is done by the methodology incorporated into the WW codes. But, in principle, it could be calculated by the special dynamic boundary layer block, which may be specially included in the model (Polnikov, 2005). This is the task for a further elaboration of the model up to the fourth generation.

## 2.4 Dis-term

In the new source function, the *Dis*-term is represented by the original and theoretically substantiated parameterization based on a special semi-phenomenological theory of the wave-turbulence interaction taking place in the upper water layer. In turn, the turbulence is supposed to be provided by different mechanisms: wave breaking, wave capping, sprinkling, shear of the upper layer currents, and so on. The basic theory was originally proposed in Polnikov (1994), and was properly specified in subsequent works (Polnikov, 2005; Polnikov and Tkalich, 2006). The present version of the term *Dis* has the form

$$Dis(\sigma, \theta, S, W) = \tilde{\gamma}(\sigma, \theta, W)\frac{\sigma^6}{g^2}S^2(\sigma, \theta) \qquad (9)$$

Note that the non-dimensional function $\tilde{\gamma}(\sigma, \theta, W)$ in (9) is found with the use of the balance ration, $In(...) \cong Dis(...)$, which is valid at the high frequency tail of the wave spectrum (for details, see the last references). This means that the dissipation-term is to some extent a consequence of the empirical input-term, and owing to this, it depends reasonably on the wind via the factor $\beta(\sigma, \theta, \mathbf{u}_*)$. Finally, for $\tilde{\gamma}(\sigma, \theta, W)$ we have the ratio

$$\tilde{\gamma}(\sigma, \theta, W) = c(\sigma, \theta, \sigma_p)\max[\beta_{dis}, \beta(\sigma, \theta, u_*, \theta_w)] \qquad (10)$$

Here, the background dissipation parameter, $\beta_{dis}$, is introduced with the aim of better correspondence to real wave dissipation under weak winds (when $\beta < \beta_{dis}$). Previously, it was found that the default value of this parameter is $\beta_{dis} = 5 \cdot 10^{-5}$. But, in principle, it is the subject of the model fitting.

In Eq. (10), the value of $\beta(\sigma, \theta, u_*, \theta_w)$ is defined by formula (8), while the non-dimensional function is given by the ratio

$$c(\sigma, \theta, \sigma_p) = C_{dis}\max\left[0, (1 - c_\sigma \cdot (\sigma_p/\sigma))\right]T(\sigma, \theta, \sigma_p), \qquad (11)$$

where the phenomenological angular function, $T(\sigma, \theta, \sigma_p)$, is introduced. It has the form

$$T(\sigma,\theta,\sigma_p) = \left\{1 + 4\frac{\sigma}{\sigma_p}\sin^2(\frac{\theta-\theta_w}{2})\right\}\max\left[1, 1-\cos(\theta-\theta_w)\right] \quad (12)$$

The dissipation cutting coefficient, $c_\sigma$, is the subject of the model fitting. The default value is $c_\sigma = 0.5$.

It is easy to see that the function $c(\sigma,\theta,\sigma_p)$, describing details of the dissipation processes in the vicinity of the peak frequency of the spectrum, $\sigma_p$, governs the dissipation rate in the whole frequency band. This corresponds to the idea of cumulative feature of the dissipation process in wind waves (Babanin and Young, 2005).

Note that in formula (10), a direct dependence of the *Dis*-term on the local wind is present. Herewith, in the case of weak winds, the condition of non-zero minimal level of wave energy loss is taking place in the $(\sigma,\theta)$–domain where $\beta(\sigma,\theta,u_*,\theta_w) < \beta_{dis}$. This fact reflects the consideration of the background dissipation processes provided by an effective viscosity for the waves in the upper layer. Such an element of parameterization of *Dis* was proposed for the first time in Polnikov (2005).

In the term *Dis*, the coefficient $C_{dis}$ is the main object of the model fitting. In this work, the parameters $c_\sigma$ and $\beta_{dis}$ are mainly used with the default values mentioned above; but, principally, they could be varied for better model fitting as well.

Hereafter, the numerical model WW, modified with the original source function replaced by the new one, is referred as the model NEW.

## 3. METHODOLOGY OF STUDYING MODEL PERFORMANCE

There are two approaches to study the numerical model performance: testing and validation. The former is based on execution of academic testing and the latter on validation of models against natural observation data. In our study, we used both approaches. The basic principles of these processes have their own specifications and it is worthwhile to mention them briefly, following Efimov and Polnikov (1991) and Komen et al. (1994).

### 3.1 Initial rules for testing the models

There are three principal features underlying the importance of the testing process. They are as follows:

(1) Possibility to reveal numerical features of the model by means of simplified consideration based on using the fully controlled wind and boundary conditions.

(2) Message comprehensibility and predictability of the testing tasks.

(3) Simple and narrow-aimed posing of the testing tasks.

There is a long list of testing tasks which could be used for the model properties evaluation (for examples, see The SWAMP group, 1985; Efimov and Polnikov, 1991; Komen et al., 1994; Polnikov, 2005). However, it is not our objective to execute all of them. At the present stage of study, we used the following list of tests.

#1. Straight fetch test (the wave development or tuning test).

#2. Swell decay test (the dissipation test).

In general, it is possible to distinguish three levels of adequacy of numerical wind wave models, which are defined by the proper choice of reference parameters used for comparison against observations (Efimov and Polnikov, 1991). But, here we restrict ourselves to the first level only as the checking of the second and third level of adequacy needs much more time and efforts. It is postponed for future studies. An example of such testing can be found in Polnikov (2005).

The first level reference parameters are the most important ones, as far they are used in test #1. They are as follows:

• non-dimensional wave energy

$$\tilde{E} = \frac{Eg^2}{W_{10}^4} \quad (\text{or } E^* = \frac{Eg^2}{u_*^4}), \text{ and} \quad (13)$$

• non-dimensional peak frequency of the wave spectrum

$$\tilde{\sigma}_p = \frac{\sigma_p W_{10}}{g} \quad (\text{or } \sigma_p^* = \frac{\sigma_p u_*}{g}), \quad (14)$$

where the dimensional energy, $E$, is calculated by the ordinary formula, $E = \iint S(\sigma,\theta)d\sigma d\theta$, and $\sigma_p$ is the peak frequency of the spectrum $S(\sigma,\theta)$.

Both values, $\tilde{E}$ and $\tilde{\sigma}_p$, estimated from simulations for a stationary stage of the wind wave field, are considered as functions of the non-dimensional fetch, $\widetilde{X} = Xg/W_{10}^2$. Numerical dependences of $\widetilde{E}(\widetilde{X})$ and $\widetilde{\sigma}_p(\widetilde{X})$, found in simulations, are to be compared with the reference empirical ratios (Komen et al., 1994):

(a) For the stable atmospheric stratification:

$$\widetilde{E}(\widetilde{X}) = 9.3 \cdot 10^{-7} \widetilde{X}^{0.77} ; \quad \widetilde{\sigma}_p(\widetilde{X}) = 12\widetilde{X}^{-0.24} \quad (15)$$

(b) For the unstable atmospheric stratification:

$$\widetilde{E}(\widetilde{X}) = 5.4 \cdot 10^{-7} \widetilde{X}^{0.94} ; \quad \widetilde{\sigma}_p(\widetilde{X}) = 14\widetilde{X}^{-0.28} \quad (16)$$

For test #2, the proper reference parameters are specified in Section 4.2 below.
On the basis of this comparison, the first tuning of the unknown coefficients in the source function, $C_{in}$, $C_{dis}$, and $C_{nl}$, is done. Thus, the purpose of these tests is to tune the model. But, here we should also say that the results of this tuning are not unequivocal (see below); and, in principle, it needs to use more complicated tasks to achieve a more sophisticated tuning. The validation process is one of these tasks.

## 3.2 Comparative validation of the models

Another approach of studying the properties of numerical models is the process of validation. But we did not carry out a formal validation procedure. Instead, a comparative validation of two models was performed. In this regard, it is worthwhile to note that the comparative validation procedure is a delicate methodological process, the main requirements of which are not well formulated till now. The proper formulations should be formalized as a series of special rules. Such formalization is planned to be done in details in a separate work. At present, as the primary initial rule, we can state that the comparative validation procedure requires meeting certain conditions. The main ones are the following:

(a) Reasonable data base including accurate and frequent wave observations;

(b) Reliable wind field given on a rather fine space-time grid for the whole period of wave observations;

(c) Properly elaborated mathematical part for a numerical model in the form (1);

(d) Certain numerical wind wave model chosen for comparison as a reference.

In our work, the last two requirements were satisfied by choosing model WW, whilst the other conditions were met by the following way.

(i) Three oceanic areas were chosen, for which the wave observation data are available for us at present: Barenz Sea, western and eastern parts of the North Atlantic (NA hereafter).

At the first stage of the validation process, we have used the one-month data (January, 2006) for 19 buoys located both in the western and in the eastern parts of NA (Fig. 1). These data have a time interval of 1 hour, which corresponds to more than 700 points of observations on each buoy.

(ii) As for the wind field, we have used a re-analysis provided by NCEP/NCAR with a spatial resolution of 1 degree both in longitude and in latitude. The time resolution for the wind was 3 hours. To exclude uncertainties associated with the boundary conditions, the simulation region was restricted by the following coordinates: 78°S–78°N latitude and 100°W–20°E longitude, and the ice covering fields were included.

The first stage of validation has been executed, basing on the above data. These calculations resulted in a sophisticated choice of the fitting coefficients, $C_{in}$, $C_{dis}$, and $C_{nl}$, found for the default values of the other fitting parameters mentioned above: $b_L$, $\beta_{dis}$, and $c_\sigma$ (see Sections 2.3 and 2.4).

At the second stage of validation, we have used the long-period historical data of the National Buoy Data Centrum (NBDC) (covering October–May period in the years 2005–2006) for 12 buoys located in the western part of NA. The wind fields and the time-space resolution were of the same features as at the first stage. Based on these data, the standard validation of both models has been done without changing any coefficients.

Finally, the control comparison of accuracy of both models was done for the region of the Barenz Sea, for which the Norwegian buoy data were used. They correspond to the 3-year wind wave observations (1990–1992) at an interval of 6 hours (totally more than 4000 points). The simulation area is shown in Fig. 2. The grid size was 1.5 degrees in longitude and 0.5 degree in latitude.

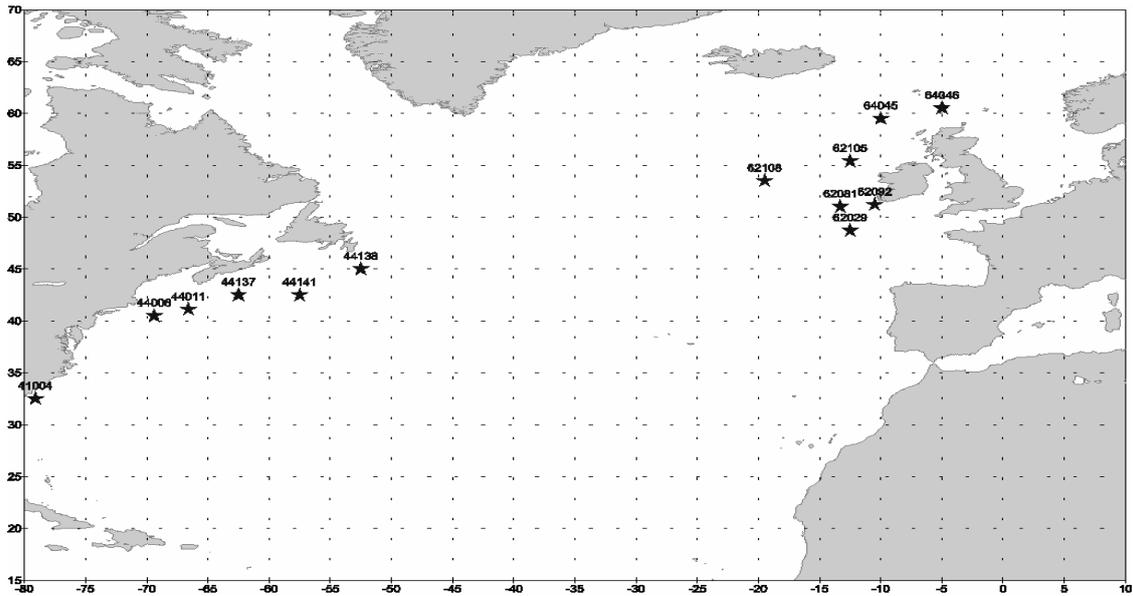

Fig. 1    Part of simulation region in the North Atlantic, showing some buoy locations.

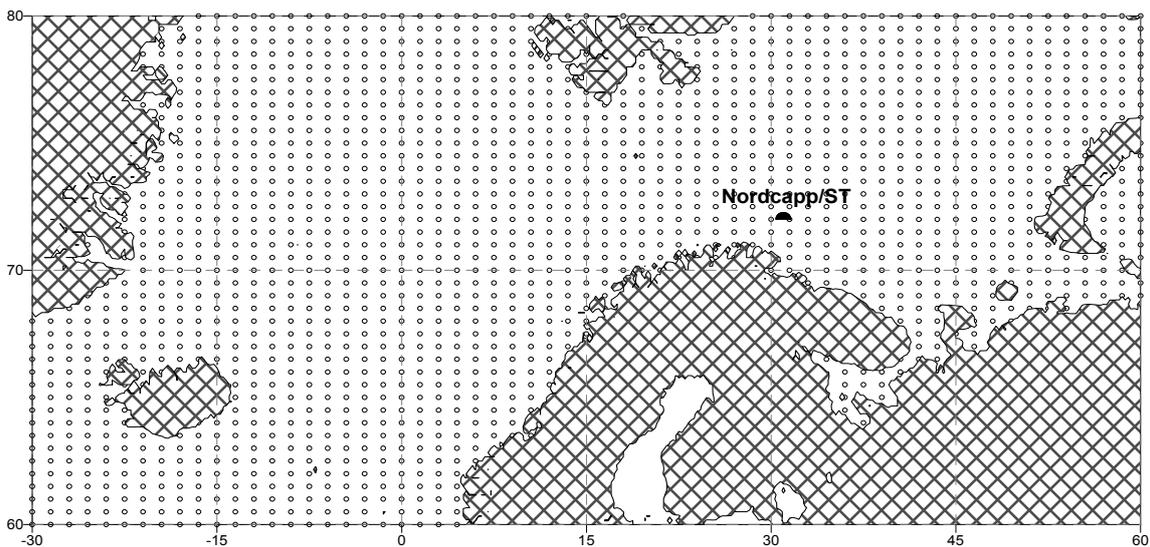

Fig. 2    Simulation area in the Barenz Sea, with position of the Norwegian buoy depicted.

Since the most reliable buoy data are related mainly to the observation of the significant wave height, $H_s$, an estimation of the simulation accuracy for this wave characteristics was done in each validation case. Similar error estimations for a peak wave period, $T_p$, and mean period, $T_m$, were executed for the two cases of long-period simulations with the aim of completeness. Detailed analysis of the latter estimations is postponed for future investigation.

### 3.3  Specification of numerical simulations

In our calculation, we have used the frequency-angle grid of the form (2), having parameters

$\sigma_0 = 2\pi \cdot 0.04$ rad, $e = 1.1$ and

$\Delta\theta = \pi/12$  (or $\Delta\theta = 15°$) \hfill (17)

with the number of frequency bins $N = 24$ and number of angle bins $M = 24$.

In the case of model testing, the spatial grid was taken in Cartesian coordinates, including 100 points in the $x$-direction and 21 points in the $y$-direction. In the case of model validation in oceanic regions, the grid was taken in spherical coordinates, with the number of grid points depending on the region (Figs. 1 and 2).

The space and time steps of calculations, $\Delta X, \Delta Y,$ and $\Delta t$, were varied in accordance with the tasks and the numerical stability conditions. In Cartesian coordinates, they varied in the limits of $\Delta X = \Delta Y = 10^3$ to $90 \times 10^3$ m and $\Delta T = 300$ to $900$ s. In spherical coordinates, they were $\Delta X = \Delta Y = 1^0$ and $\Delta T = 1200$ s. Every time, an initial spectrum was taken in the frame of WW codes.

### 3.4 Statistical measures of the validation errors

To assess the accuracy of simulating a time-series of a certain wave parameter $P(t)$, we have used the following error estimates:

(a) the root-mean-square error, $\delta P$, given by the formula

$$\delta P = \left( \frac{1}{N_{obs}} \sum_{n=1}^{N_{obs}} \left( P_{num}(n) - P_{obs}(n) \right)^2 \right)^{1/2} \quad (18)$$

and

(b) the relative root-mean-square error, $\rho P$, defined as

$$\rho P = \left( \frac{1}{N_{obs}} \sum_{n=1}^{N_{obs}} \left( \frac{P_{num}(n) - P_{obs}(n)}{P_{obs}(n)} \right)^2 \right)^{1/2} \quad (19)$$

Here $N_{obs}$ is the total number of observation points taken into consideration, and self-explanatory subscripts are used.

In addition to this, the following arithmetic error was used for analysis:

$$\alpha P = \left( \frac{1}{N_{obs}} \sum_{n=1}^{N_{obs}} \left( P_{num}(n) - P_{obs}(n) \right) \right) \quad (20)$$

The first two errors describe statistical scattering of the simulating results (or the errors of the input fields, like a wind), whilst the latter one represents the mean shift of numerical results with respect to observations.

There are several other statistical characteristics which could be useful for assessment of a numerical model quality (correlation coefficient, probability function, and so on; for examples, see Tolman et al., 2002). But at the present stage of validation, they are omitted for the sake of clarity of the primary analysis of the results presented below.

## 4. RESULTS OF MODEL TESTING

### 4.1 Straight fetch test

Pose of the task. Spatially homogeneous and invariable in time wind, $W(x,t) = W_{10} = const$, is blowing normally to a very long straight shore line. Initial conditions are given by a homogeneous wave field with a wave spectrum of small intensity. Boundary conditions are constant in time and correspond to the initial wave state.

The purpose of the test is to check correspondence of the wind wave growing curves, $\widetilde{E}(\widetilde{X})$, $\widetilde{\sigma}_p(\widetilde{X})$, provided by the model, with the reference empirical growing curves for the stationary state of developed wind waves given by ratios (15) and (16).

Since the results of this test are typical and well predicted, here we show only some examples of the testing results of the model NEW for different wind values, $W_{10} = 10 - 30$ m/s. They are presented in Figs. 3–5, for values of $C_{nl} = 9 \cdot 10^7$, $C_{in} = 0.4$, $C_{dis} = 60$, and the default values for the other fitting parameters (see Sections 2.2–2.4). The proper results for the original WW are presented, for example, in Tolman and Chalikov (1996).

From Figs. 3–5 one can see that curves of $\widetilde{E}(\widetilde{X})$ and $\widetilde{\sigma}_p(\widetilde{X})$ of the modified model are in good agreement with empirical ratios (15) and (16). It shows a good degree of tuning of the model, at least adequate at the first level.

For completeness of treating the results shown, it is worthwhile to note the following.

First, the jumps between simulation curves, presented in Figs. 3–5, are provided by a change of the spatial step, $\Delta X$ and $\Delta Y$. This increase of the spatial step by ten times was done in our calculations with the aim to cover a large range of non-dimensional fetches, $\widetilde{X}$, for a fixed wind velocity, $W_{10}$[1]. Such a jump for $\widetilde{E}(\widetilde{X})$ and $\widetilde{\sigma}_p(\widetilde{X})$ is a typical feature of any numerical scheme used in the model, resulting in an inevitable dependence of numerical errors on the value of spatial step. Usually, these errors are magnified at the points located near the shoreline.[2]

---

[1] For $\Delta X = 10^3$ m and $W_{10} = 10$ m/s, the range of the non-dimensional fetch $\widetilde{X}$ was $10^2 \leq \widetilde{X} \leq 10^4$, and for $\Delta X = 10^4$, it was $10^3 \leq \widetilde{X} \leq 10^5$.

[2] This point is mainly related to the mathematical part of the model, which is not discussed here. Evidently, it should be elaborated further in more details.

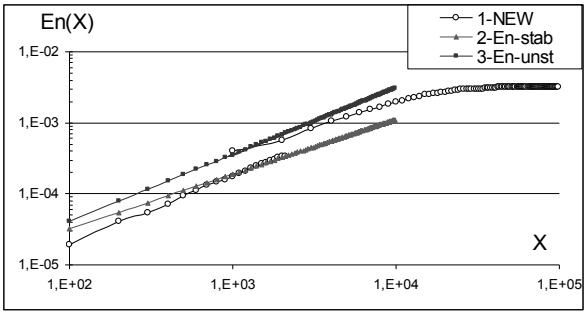

Fig. 3  Dependence of non-dimensional peak energy on non-dimensional fetch, $\widetilde{E}(\widetilde{X})$, for $W_{10} = 10$ m/s : 1–model NEW; 2–stable stratification; 3–unstable stratification.

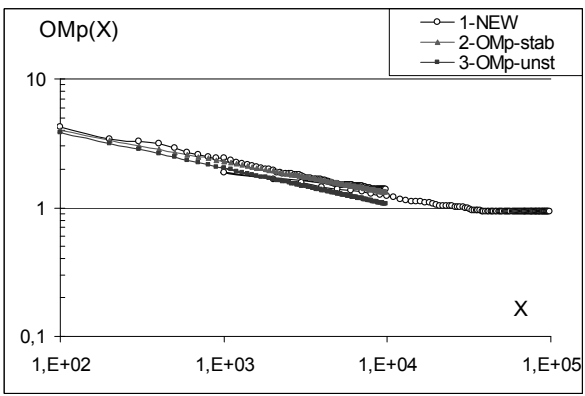

Fig. 4  Dependence of non-dimensional peak frequency on non-dimensional fetch, $\widetilde{\sigma}_p(\widetilde{X})$. For legend, see Fig. 3.

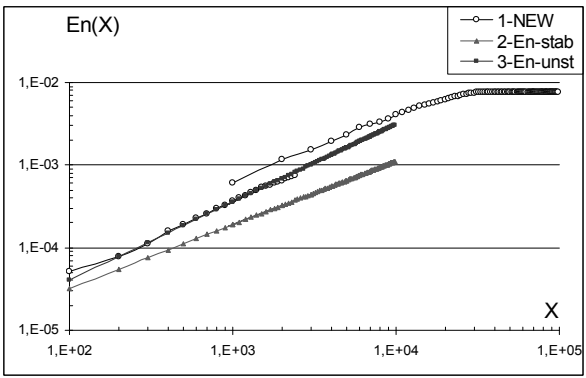

Fig. 5  Dependence of non-dimensional peak energy on non-dimensional fetch, $\widetilde{E}(\widetilde{X})$ for $W_{10} = 30$ m/s. For legend, see Fig. 3.

Additionally, in our presentation of the reference parameters (see formulas (13) and (14)), the location of the curve $\widetilde{E}(\widetilde{X})$ is shifted for the same fetch, $\widetilde{X}$, while changing values of $W_{10}$. This result is also typical (see Komen et al, 1994), taking into account the dependence of friction velocity, $u_*$, on $W_{10}$, realized in WW. The shifting effect is greatly reduced, if one represents the non-dimensional parameters in terms of $u_*$ (i.e., the dependencies $E^*(X^*)$ and $\sigma^*_p(X^*)$. But, this artificial effect is not important enough to warrant further consideration (for details, refer to Komen et al., 1994; Tolman and Chalikov, 1996). Second, it should be taken into account that the empirical dependences (15) and (16) are valid for non-dimensional fetches of the range $10^2 \leq \widetilde{X} \leq 10^4$ with the errors of the order of 10–15% (Komen et al., 1994). This natural scattering feature of empirical data provides for a possibility to fit a lot of different models to the dependences (15) and (16) with the same accuracy.

Third (and it is the most important), a good correspondence of numerical with the empirical dependences $\widetilde{E}(\widetilde{X})$ and $\widetilde{\sigma}_p(\widetilde{X})$ does not imply an unequivocal choice of the fitting parameters. Root-mean-square errors of the order of 10–15% can be achieved for a continuum of values of the fitting parameters, such as $C_{in}$, $C_{dis}$, $C_{nl}$, and the others mentioned in Sections 2.2–2.4. This result is provided for the simplified meteorological conditions used in the testing task. The sophisticated fitting of the model could be achieved only by means of the model validation against observations executed for a rather long period of wave evolution under well controlled but varying meteorological conditions. This point will be discussed in details in Section 5 below.

### 4.2 Swell decay test

A forcing wind with a fixed value is present in the first part of the testing area: $W(X) = W_{10}$ at points $0 \leq X \leq X_m$. In the second part of the area, the wind is absent: $W(X) = 0$ at $X_m < X \leq 3X_m$. The initial wave state and boundary conditions are typical (see test #1 above).

The numerical evolution is continued for the period $T$ for a full development of waves at the fetch $X = X_m$ and for the decaying swell field to reach a stable state in the second part of the testing area. The corresponding value of the non-dimensional time, $\widetilde{T} = Tg/W_{10}$, should be about several times of $10^5$.

The aim of the test is to reveal quantitative features of the swell decay process, starting from the fully developed sea with different peak frequencies, $f_{sw} = f_p(X_m)$. The latter is

Engineering Applications of Computational Fluid Mechanics Vol. 2, No. 4 (2008)

considered as the principal initial characteristic of the swell. (Here we take into account that the initial intensity of the swell is mainly provided by $f_{sw}$).

To achieve the stated aim, different values of $W_{10}$, $X_m$, and $T$ should be taken into consideration. In our calculations, for a wind $W_{10} = 10$ m/s, we took $\Delta X = 10$ km, $X_m = 240$ km and $T = 48$ h; and for $W_{10} = 20$ m/s, we used $\Delta X = 40$ km, $X_m = 760$ km and $T = 72$ h.

In the second part of the area, the following reference parameters are checked:

- the relative energy lost parameter given by the ratio

$$Ren(X) = E(X - X_m)/E(X_m) \qquad (21)$$

- the relative frequency shift parameter defined as

$$Rf_p(X) = f_p(X - X_m)/f_p(X_m) \qquad (22)$$

As far as there are no widely recognized empirical dependences $Ren(X)$ and $Rf_p(X)$, the ones found are evaluated at the expert level, only. The latter means a qualitative physical analysis of the numerical results (see below).

Results of our simulation are shown in Figs. 6 and 7, representing the swell decay process for values $W_{10} = 10$ and $20$ m/s. The correspondent values of the initial swell frequency, $f_{sw}$, are 0.18 Hz and 0.085 Hz, respectively.

From these figures one can draw the following conclusions.

(1) The rate of swell energy dissipation depends strongly on the initial peak frequency of swell, $f_{sw}$. This rate is quickly going down with the distance of swell propagation (Fig. 6).

(2) The swell dissipation rate for the model NEW is faster than for WW (Fig. 6).

(3) The rate of peak frequency shifting to lower values, provided by the nonlinear interaction between waves, depends strongly on the initial value of peak frequency, $f_{sw}$ (Fig. 7). The greater the $f_{sw}$, the greater the rate of frequency shifting. This is well understood, taking into account formula (6) for the nonlinear evolution term.

(4) The model NEW has practically the same rate of the peak frequency shifting, in contrast to the rate of the relative energy loss (Fig. 7).

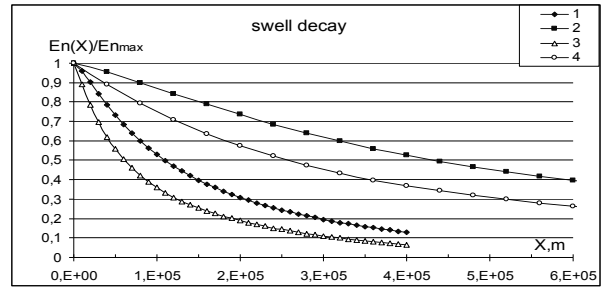

Fig. 6  Dependence $Ren(X)$ for two values of initial peak frequency of swell: 1, 2–original model WW; 3, 4–model NEW; 1, 3–$f_{sw}$=0.18 Hz; 2, 4–$f_{sw}$=0.085 Hz.

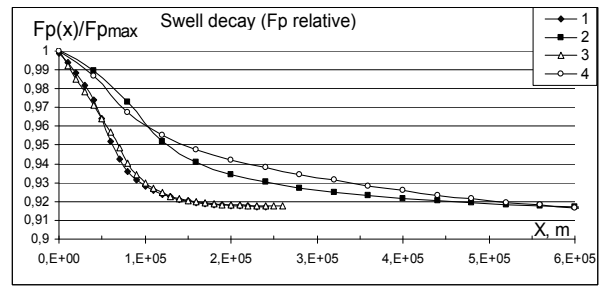

Fig. 7  Dependence $Rf_p(X)$ for two values of initial peak frequency of swell. For legend, see Fig. 6.

This test is very instructive in the physical aspect. In fact, from the results obtained, one can infer the following.

First, from conclusion (2), one can state that the new dissipation term is more intensive than the one used in the original model WW.

Second, from conclusion (4), one can state that there exists a very close similarity between the nonlinear terms in both models.

Third, from the previous two inferences, one could state that the main qualitative difference of the numerical results obtained for these two models is mainly provided by the new parameterization of the *Dis*-term. Herewith, we note that though the new parameterization of the *In*-term has a feature of additional background dissipation, in this test it is too small to play any remarkable role, especially at the initial stage of swell decay.

As one could see later, the last inference is of the most importance for understanding and handling the difference between these models, which will be found during validation.

## 5. RESULTS OF COMPARATIVE VALIDATION OF THE MODELS WW AND NEW

### 5.1 One-Month simulations in the North Atlantic

After several runs of the model NEW, intended for a sophisticated choice of the fitting coefficients $C_{in}$, $C_{dis}$, and $C_{nl}$, we have found that the best results (i.e., minimum errors $\delta H_s$ for the majority of buoys) are obtained for the following values:

$$C_{nl} = 9 \cdot 10^7, \ C_{in} = 0.4; \ C_{dis} = 70, \text{ and } c_\sigma = 0.7 \quad (23)$$

with the default values of the other fitting parameters.

A typical time history of the significant wave height, $H_s(t)$, obtained in these simulations is shown in Fig. 8 for buoy 41001 chosen as an example. From this figure, in particular, one can see that the model NEW follows the extreme values of real waves better than it is done by the model WW. Visual analysis of all proper curves has shown that this feature of the model NEW is typical for the majority of buoys taken into consideration. More detailed and quantitative analysis needs to be conducted using the statistical procedures based on the error measures described above in Section 3.4.

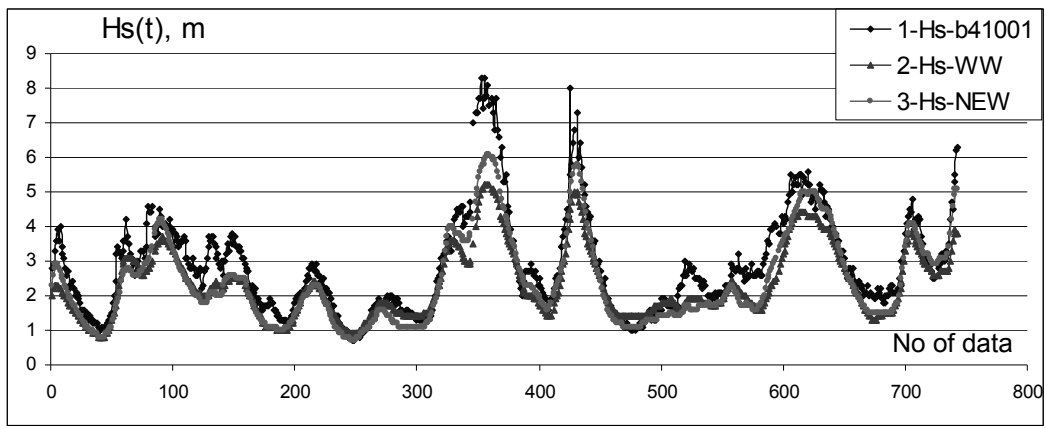

Fig. 8   Time history of the observed and simulated wave heights, $H_s(t)$, on buoy 41001 for January 2006. 1–wave heights measured on the buoy; 2–wave heights simulated by the model WW; 3–wave heights simulated by the model NEW.

Table 1  Root-mean-square errors of simulations in the eastern part of the North Atlantic (NA).

| Eastern NA, No. of buoy | Model WW | | Model NEW | | $\frac{(\delta H_s)_{WW}}{(\delta H_s)_{NEW}}$ |
|---|---|---|---|---|---|
| | $\delta H_s$, m | $\rho H_s$, % | $\delta H_s$, m | $\rho H_s$, % | |
| 62029 | 0.57 | 14 | 0.54 | 13 | 1.05 |
| 62081 | 0.67 | 15 | 0.56 | 13 | 1.20 |
| 62090 | 0.66 | 14 | 0.57 | 14 | 1.16 |
| 62092 | 0.58 | 14 | 0.53 | 14 | 1.09 |
| 62105 | 0.79 | 18 | 0.68 | 15 | 1.16 |
| 62108 | 0.99 | 15 | 0.84 | 13 | 1.18 |
| 64045 | 0.71 | 12 | 0.61 | 12 | 1.16 |
| 64046 | 0.72 | 15 | 0.76 | 15 | 0.95 |

Table 2  Root-mean-square errors of simulations in the western part of the North Atlantic (NA).

| Western NA, No. of buoy | Model WW | | Model NEW | | $\frac{(\delta H_s)_{WW}}{(\delta H_s)_{NEW}}$ |
|---|---|---|---|---|---|
| | $\delta H_s$, m | $\rho H_s$, % | $\delta H_s$, m | $\rho H_s$, % | |
| 41001 | 0.81 | 22 | 0.66 | 20 | 1.23 |
| 41002 | 0.52 | 18 | 0.47 | 18 | 1.11 |
| 44004 | 0.82 | 25 | 0.68 | 26 | 1.21 |
| 44008 | 0.83 | 27 | 0.61 | 24 | 1.36 |
| 44011 | 0.82 | 23 | 0.55 | 18 | 1.49 |
| 44137 | 0.58 | 19 | 0.51 | 17 | 1.14 |
| 44138 | 0.70 | 19 | 0.74 | 19 | 0.95 |
| 44139 | 0.63 | 19 | 0.69 | 20 | 0.91 |
| 44140 | 0.78 | 19 | 0.80 | 19 | 0.97 |
| 44141 | 0.64 | 20 | 0.68 | 20 | 0.94 |
| 44142 | 0.81 | 27 | 0.48 | 18 | 1.69 |

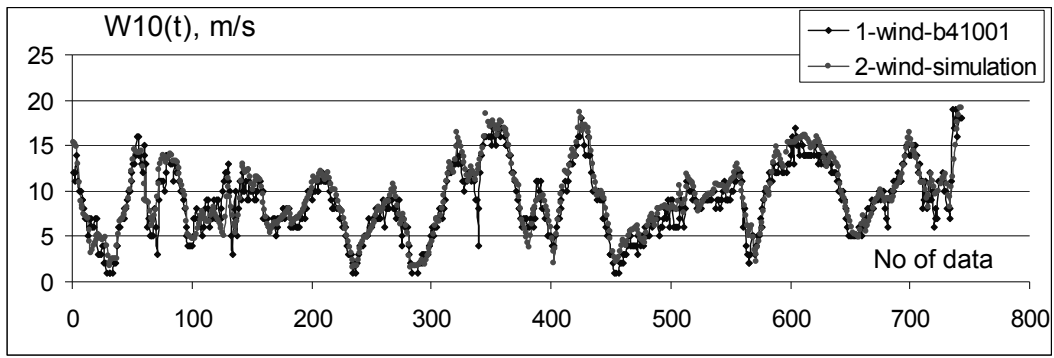

Fig. 9  Time history of the observation and simulation wind, $W_{10}(t)$, on buoy 41001 for January 2006.
1–wind measured on the buoy; 2–wind used in the model simulation.

At this stage of validation, the properly estimated errors have been found for the significant wave height, $H_s$, only. They are presented in Tables 1 and 2 for two parts of NA respectively. For a quick general (visual) evaluation of the results, we have shaded cells corresponding to the cases when the model NEW has a loss of accuracy.

The analysis of these results leads to the following conclusions.

First, the accuracy of the model NEW is better than the model WW for more than 70% of buoys considered.

Second, discrepancy of the r.m.s. errors for both models is remarkable. The typical win accuracy of the model NEW is of the order of 15–20%, however, sometimes it can reach 70% (buoy 44142).

Third, the relative error, $\rho H_s$, calculated by taking into account each point of observations, is comparable (15–27%). It has a tendency of error reduction in the model NEW, but this is not so obvious.

Basing on the above results, we should note that in the present statistical form of consideration, the relative error $\rho H_s$ is not so sensitive to the specificity of the model, as could be expected. It seems that the effect of increased sensitivity of $\rho H_s$ could arise, if we introduce the lower limit of the wave heights into the procedure of error estimation. For example, the proper error estimations could be done, if one excludes the time-series points $H_s(t)$ for values below 2m. But the effect of the introduction of limiting values for $H_s$ (or for $T_p$) is not so evident, therefore this issue should be further studied later.

In this connection, it is worthwhile to mention about an accuracy of the input wind. The proper time history for $W_{10}(t)$ is shown in Fig. 9.

From first sight, the correspondence between the simulation wind and the observed wind seems to be rather good. But direct calculations of the errors $\delta W_{10}$ and $\rho W_{10}$, made, for example, for buoy 41001, give the values

$$\delta W_{10} = 1.56 \text{ m/s} \quad \text{and} \quad \rho W_{10} = 32\% \tag{24}$$

The first value is more or less reasonable, taking into account that the input wind is calculated by the re-analysis covering a very large domain. But the last value in (24), $\rho W_{10}$, seems to be fairly large with respect to the corresponding relative error $\rho H_s$ (Table 1). Due to an arbitrary choice of the buoy considered, one can expect that such a mismatch between values of $\rho H_s$ and $\rho W_{10}$ is typical for the present consideration, which in turn entails a proper explanation.

This mismatch of values for $\rho W_{10}$ and $\rho H_s$ leads to a pose of the following new task: how to treat the present inconsistency between these errors. To solve this problem, first of all, it requires statistics of a large amount of error values. Such statistics will be presented in Table 3 below. Besides, physically it is reasonable to introduce the lower limiting values for wind, $W_{10}$, and wave heights, $H_s$, which restrict the proper time-series points involved in the procedure of error estimation. That way, one could find a physically reasonable, unequivocal interrelation between errors $\rho W_{10}$ and $\rho H_s$. If it is found, this relation permits a proper physical treatment of the errors and clear the way to numerical modeling improvements. This work is postponed to a future investigation.

### 5.2  Long-period simulations in the western part of the North Atlantic

Simulation results for the second stage of validation are very similar to the ones presented above. The proper errors are shown in Table 3,

where the shaded cells correspond to the cases of less accuracy of $H_s$ for the model NEW.

From this table, in general, one can see a reasonable advantage of the new model with respect to WW, in the aspect of simulation accuracy for the wave heights, which is defined by the values of r.m.s. error $\delta H_s$. The improvement in accuracy is in the range of 1.1–1.5 times.

More detailed analysis results in the following. Arithmetic errors for WW are usually greater than the corresponding values for the model NEW. From Table 3 it is seen that the model WW always underestimates the wave heights, $H_s$, whilst the model NEW has more symmetrical and smaller arithmetic errors. These facts allow us to conclude that the model NEW (and the new source function, consequently) has apparently better physical grounds.

In the aspect of accuracy for the wave periods, $T_m$ and $T_p$, we should confess that the model NEW has less accuracy of calculation for the mean wave period, $T_m$, but, it has practically the same (or even better) accuracy for the peak wave period, $T_p$ (Table 3).

Regarding the wave periods, we should note a very specific feature, consisting in the fact that both models show a certain overestimation of the mean wave period, $T_m$, whilst the peak period, $T_p$, is always underestimated. The most probable reason for such a behavior of the models could be attributed to an insufficient accuracy in the calculation of the 2D-shape of the wave spectrum, $S(\sigma,\theta)$, for both models.

Regarding the mismatch of the mean wave period, one may additionally suppose that this effect could be related to the methodology of quantitative estimation of $T_m$, realized in the buoy equipment. The systematic error could be provided by the automatic calculation of the 1D-spectrum of wind waves, $S(\sigma)$, currently (hourly) done with the aim of attaining estimation for $T_m$.

Here we should note that in accordance with the definition

$$T_m = \frac{2\pi \int \sigma^{-1} S(\sigma) d\sigma}{\int S(\sigma) d\sigma}, \qquad (25)$$

a twice more accurate estimation of the spectrum function, $S(\sigma)$, is needed to meet the proper requirements for an accurate evaluation of $T_m$. It is well known that an accurate estimation of $S(\sigma)$ to be done correctly in a quantitative approach is not a simple task (Bendat and Piersol, 1971). So, this question needs to be examined more thoroughly and the documentation of the buoys construction be studied closely.

Thus, the definite conclusion about superiority of one model against the other cannot be drawn at present. Nevertheless, in principle, it could be done later after the proper criteria have been formulated. This point is posed here, and we plan to solve it in our future work.

### 5.3 Three-year simulation in the Barenz Sea

This case is presented here only for examining the quality of fitting of the model NEW, based on the available wind field and wave observation base (a history of this issue is presented in Polnikov et al., 2008). Statistics of the proper errors are presented in Table 4.[3]

As one can see in this case study, the model NEW has better accuracy for the wave heights, and yet inaccuracy for the mean wave periods. The values of the relative errors $\rho H_s$ are rather large, whilst the errors $\rho T_m$, in contrast, are quite reasonable and smaller than the ones presented in Table 3. Such a result for $\rho H_s$ is possibly due to the (bad) quality of the wind field. Regarding the small errors $\rho T_m$, one may say that this effect is possibly due to the particular engineering design of the buoy construction (here we mean the automatic estimation of $T_m$).

Finally, we note that the first part of the above conclusions leads again to the necessity of a detailed study of the impact of wind field uncertainty on the accuracy of wave simulations. We plan to do it in a future project.

### 5.4 Speed of calculation

By using the numerical procedure PROFILE, we have checked the speed of calculation, realized while executing all main numerical subroutines used in the models. In terms of computer processing time, the time distributions among the main subroutines are shown for both models in Table 5. These distributions correspond to the case of executing the task of 24-hour simulation of the wave evolution in the whole Atlantic.

---

[3] Unfortunately, in this case, the wind $W_{10}$ and the wave peak period $T_p$ were not provided by the Norwegian buoy.

Engineering Applications of Computational Fluid Mechanics Vol. 2, No. 4 (2008)Table 3  Consolidated input and output errors for the 8-month simulations in the western part of the North Atlantic (NA).

| No. of buoy/model | $\delta W_{10}$, m/s | $\rho W_{10}$, % | $\delta H_s$, m | $\rho H_s$, % | $\delta T_m$, s | $\rho T_m$, % | $\delta T_p$, s | $\rho T_p$, % | $\alpha W_{10}$, m/s | $\alpha H_s$, m | $\alpha T_m$, s | $\alpha T_p$, s | $\dfrac{(\delta H_s)_{ww}}{(\delta H_s)_{new}}$ | $\dfrac{(\alpha H_s)_{ww}}{(\alpha H_s)_{new}}$ |
|---|---|---|---|---|---|---|---|---|---|---|---|---|---|---|
| 41001/WW | 2.01 | 40 | 0.68 | 22 | 0.93 | 17 | 2.02 | 24 | 0.58 | -0.45 | 0.46 | -1.32 | 1.42 | 2.04 |
| /NEW |  |  | 0.48 | 18 | 1.23 | 22 | 2.13 | 30 |  | -0.22 | 0.79 | -0.88 |  |  |
| 41002/WW | 1.77 | 48 | 0.48 | 19 | 1.20 | 22 | 2.01 | 27 | 0.25 | -0.23 | 0.78 | -1.05 | 1.09 | 7.67 |
| /NEW |  |  | 0.44 | 20 | 1.58 | 30 | 2.22 | 35 |  | -0.3 | 1.11 | -0.57 |  |  |
| 41004/WW | 2.54 | 36 | 0.97 | 51 | 1.33 | 31 | 2.40 | 36 | -1.48 | -0.97 | 0.63 | -1.26 | 1.52 | 2.06 |
| /NEW |  |  | 0.64 | 36 | 1.36 | 32 | 2.38 | 38 |  | -0.47 | 0.73 | -1.10 |  |  |
| 41010/WW | 1.24 | 32 | 0.40 | 19 | 1.61 | 33 | 2.06 | 29 | 0.09 | -0.19 | 1.25 | -0.88 | 1.08 | 2.37 |
| /NEW |  |  | 0.37 | 20 | 2.07 | 43 | 2.34 | 41 |  | -0.08 | 1.69 | -0.21 |  |  |
| 41025/WW | 2.18 | 50 | 0.44 | 24 | 1.47 | 30 | 2.23 | 30 | 0.47 | -0.05 | 1.09 | -1.16 | 0.81 | 0.29 |
| /NEW |  |  | 0.54 | 31 | 1.82 | 38 | 2.23 | 35 |  | 0.17 | 1.48 | -0.57 |  |  |
| 41040/WW | 0.91 | 20 | 0.22 | 10 | 1.78 | 30 | 1.87 | 18 | 0.08 | -0.10 | 1.64 | -0.90 | 0.88 | 1.43 |
| /NEW |  |  | 0.25 | 11 | 2.11 | 35 | 1.96 | 22 |  | -0.07 | 1.90 | -0.53 |  |  |
| 41041/WW | 0.96 | 22 | 0.20 | 09 | 1.92 | 32 | 2.22 | 21 | 0.17 | -0.06 | 1.78 | -0.90 | 0.87 | 1.50 |
| /NEW |  |  | 0.23 | 10 | 2.26 | 38 | 2.20 | 24 |  | 0.04 | 2.06 | -0.54 |  |  |
| 44004/WW | 1.91 | 40 | 0.72 | 24 | 1.13 | 21 | 1.96 | 24 | 0.16 | -0.38 | 0.52 | -1.34 | 1.26 | 9.5 |
| /NEW |  |  | 0.57 | 24 | 1.32 | 25 | 1.88 | 26 |  | -0.04 | 0.82 | -1.00 |  |  |
| 44005/WW | 2.28 | 59 | 0.58 | 25 | 1.44 | 30 | 2.27 | 38 | 1.19 | -0.30 | 0.84 | -0.84 | 1.29 | 10.0 |
| /NEW |  |  | 0.45 | 27 | 1.78 | 37 | 2.24 | 43 |  | 0.03 | 1.37 | -0.19 |  |  |
| 44008/WW | 2.35 | 51 | 0.70 | 25 | 1.11 | 21 | 2.01 | 26 | 0.69 | -0.43 | 0.60 | -1.31 | 1.4 | 4.78 |
| /NEW |  |  | 0.50 | 21 | 1.30 | 25 | 1.88 | 28 |  | -0.09 | 0.91 | -0.90 |  |  |
| 44014/WW | Bad information | Bad information | 0.49 | 26 | 1.17 | 21 | 2.46 | 31 | Bad information | -0.27 | 0.46 | -1.66 | 1.4 | 2.45 |
| /NEW |  |  | 0.35 | 23 | 1.27 | 24 | 2.37 | 33 |  | -0.11 | 0.83 | -1.23 |  |  |
| 44018/WW | 3.01 | 43 | 0.44 | 23 | 1.14 | 22 | 2.03 | 25 | 0.61 | -0.03 | 0.72 | -1.25 | 0.80 | 0.10 |
| /NEW |  |  | 0.55 | 33 | 1.37 | 28 | 1.88 | 28 |  | 0.29 | 1.03 | -0.85 |  |  |

479

Table 4 R.M.S. errors of simulations in the region of the Barenz Sea.

| Model | $\Delta H_s$, m | $\rho H_s$, % | $\Delta T_m$, s | $\rho T_m$, % |
|---|---|---|---|---|
| WW | 0.78 | 32 | 0.89 | 16 |
| NEW | 0.68 | 27 | 0.96 | 17 |

Table 5 Distribution of computer processing time utilized by the two versions of WW.

| Model | Name of procedure (*explanation*) | Time, s | Time, % |
|---|---|---|---|
| Original WW | w3snl1md_w3snl1 (*Nl-term calculation*) | 123.41 | 27.06 |
| | w3pro3md_w3xyp3 (*space propagation scheme*) | 87.01 | 19.08 |
| | w3uqckmd_w3qck3 (*time evolution scheme-3*) | 68.58 | 15.04 |
| | w3iogomd_w3outg (*output of results*) | 37.73 | 8.27 |
| | w3src2md_w3sin2 (*In-term calculation*) | 21.99 | 4.82 |
| | w3uqckmd_w3qck1 (*time evolution scheme-1*) | 17.66 | 3.87 |
| | w3srcemd_w3srce (*integration subroutine*) | 13.29 | 2.91 |
| | w3src2md_w3sds2 (*Dis-term calculation*) | 2.75 | 0.60 |
| | others | … | … |
| | **All procedures** | **455.9** | **100** |
| Modified WW | w3pro3md_w3xyp3 | 89.72 | 22.52 |
| | w3uqckmd_w3qck3 | 71.29 | 17.88 |
| | w3snl1md_w3snl1 (*Nl-term*) | 70.97 | 17.80 |
| | w3iogomd_w3outg | 38.60 | 9.68 |
| | w3uqckmd_w3qck1 | 17.97 | 4.51 |
| | w3srcemd_w3srce | 12.15 | 3.05 |
| | w3src2md_w3sds2 (*Dis-term*) | 7.68 | 1.93 |
| | w3src2md_w3sin2 (*In-term*) | 6.04 | 1.52 |
| | others | … | … |
| | **All procedures** | **398.8** | **100** |

From this table one can see that in the model NEW, the speed of calculation of the nonlinear term is 1.73 times faster than that of the original WW. It leads to a reduction of calculation time of about 60 seconds, resulting in a 15% reduction of the total calculation time. The acceleration effect is provided by using the fast DIA approximation mentioned above in Section 2.2. An additional 3% reduction of time is gained owing to the new parameterization of the input term. But, in turn, the new approximation of *Dis*-term results in a loss of calculation speed of about 2%. Nevertheless, as we said above in conclusions given in Section 4.2, this parameterization leads to better accuracy of the model NEW, because of the fact that the physics of the *NL*-term and *In*-term in both models is very similar.

## 6. CONCLUSIONS

The new source function was tested and validated by incorporating it into the mathematical shell of the reference model WW. Results of test #1 are typical for any modern numerical model and are used for the primary tuning. But test #2 is more physical. It testifies the specific properties of the proposed dissipation term. The real performance of the new model was checked during the comparative validation process, which was executed in three steps differing both by duration of simulations and by regions of the world oceans. In general, we may state that both models have rather high performance, making them the best among the present models, in regard of the results of WW's validation represented in Tolman et al. (2002). The comparative validation has shown a real advantage of the model NEW with respect to the original WW, especially in the accuracy of the wave heights calculation. It gives the advantages of reducing the simulation errors for significant wave height, $H_s$, by 10 to 50% and increasing the speed of calculation by 15%.

An analysis of the curves such as those presented in Fig. 8 shows that the largest percentage of the r.m.s. error is contributed by the time-series points with extreme values of the wave heights and by the points corresponding to the phases of wave dissipation, at which the wave intensity is going down. Both of these features are controlled by the dissipation mechanism of wave evolution. Based on these grounds, we conclude that the dissipation term is parameterized more efficiently in the new model than in the original WW. This property of the model NEW is very important in its application in risk assessment.

In our study, the relative r.m.s. error, $\rho H_s$, is introduced as one of the most instructive measures for estimating the accuracy of the wave heights simulations. We found that this parameter has mean values of the order of 12–35% for both models. It is reasonable to suppose that the magnitudes of $\rho H_s$ should depend on the value of inaccuracy for the wind field used as an input. In view of this, a new task is posed: searching for a quantitative relation between the errors for waves, $\rho H_s$, and the errors of input wind, $\rho W_{10}$. This relation is predictable, taking into account the experimental ratios like (15) and (16). The study is planned to be done in a future project.

There are several other tasks related to the further validation and elaboration of the numerical wind wave models. One of them is to establish a certain upper limits of inaccuracy for wind field and for wave observations, which are required for further progress in wind wave modeling. Estimation of these limits is a priority research endeavour.

At present, it seems that the main requirement, which defines the limits of further elaboration of the numerical wind wave models, consists in using the wind field having inaccuracy below the limits mentioned above.

## ACKNOWLEDGEMENTS

The work was done at CPTEC/INPE, while Prof. V. Polnikov was a visiting scientist under the support of the FAPESP, projects #2006/56101-6.